\documentstyle[preprint,tighten,prl,aps,psfig]{revtex}
\begin{document}
\draft
\title{Aging Relation for Ising Spin Glasses} 
\author{Yukiyasu Ozeki}
\address{Department of Physics, Tokyo Institute of Technology, \\ 
Oh-okayama, Meguro-ku, Tokyo 152, Japan}
\date{\today}
\maketitle
\begin{abstract}
We derive a rigorous dynamical relation on aging phenomena
--- the aging relation --- for Ising spin glasses
using the method of gauge transformation.
The waiting-time dependence of the auto-correlation function
in the zero-field-cooling process is equivalent with that
in the field-quenching process.
There is no aging on the Nishimori line;
this reveals arguments for dynamical properties of 
the Griffiths phase and the mixed phase.
The present method can be applied to other gauge-symmetric models
such as the XY gauge glass.
\end{abstract}
Slow dynamics is a key concept to study complex systems
such as spin glasses, structure glasses, polymers, superconductors,
neural networks, and so on.
It is one of peculiar properties to characterize the spin glass (SG) phase
\cite{BindeY86,FischH91,Takaya95}.
In the mean field theory, the non-ergodicity below 
the freezing temperature is significant to understand the nature of 
spin glasses \cite{SompoZ81,MezaPV87}.
The aging phenomenon \cite{Struik78}
is a typical realization of slow dynamics especially in the SG
\cite{LundEA83,NordEA87,VincHO92,LeflEA94}.
It was first observed for a metallic SG material $\rm CuMn$ 
in the zero-field-cooling (ZFC) process \cite{LundEA83}. 
The relaxation of the isothermal remanent magnetization 
depends on the waiting time in which 
the sample is kept at constant temperature prior to the field application.
Similar waiting time dependence was observed
for other metallic SG materials and short-range SG materials
({\it e.g.} 
$\rm AgMn$, $\rm AuFe$, $\rm Fe_{\it x}Mn_{\it 1-x}TiO_3$, 
$\rm CdCr_{\it x} In_{\it 1-x} S_4,\cdots$), 
and on the thermo-remanent magnetization 
in the field-cooling (FC) process
\cite{NordEA87,VincHO92,LeflEA94}.

Several attempts have been made theoretically to explain 
the aging by phenomenological arguments 
\cite{KoperH88,FisheH88,SibanH89}.
Cugliandolo and Kurchan \cite{CugliK94} investigated the aging 
for the Sherrington-Kirkpatrick (SK) model analytically.
They examined the auto-correlation function for the nonequilibrium process
from a random state (ZFC from $T=\infty$).
Rieger \cite{Rieger93} investigated the aging 
in the 3D $\pm J$ Ising model by the Monte Carlo simulation.
He measured the waiting-time dependence of the auto-correlation function
from the all-up state (field quenching; see below).

In the present article,
we derive a rigorous dynamical relation on aging phenomena
for Ising SG models
between two nonequilibrium processes shown in Fig.\ 1. 
The auto-correlation functions 
with a waiting time $t_{\rm w}$
for these processes have a relation,
\begin{eqnarray}
\left[\langle S_i (t_{\rm w}) S_i (t+t_{\rm w})
\rangle_K^{\bf F}\right]_{\rm c} =
\left[\langle S_i (t_{\rm w}) S_i (t+t_{\rm w})
\rangle_K^{K_p}\right]_{\rm c},
\label{eq:AR}
\end{eqnarray}
where $K=J/k_{\rm B} T$ is the inverse temperature of the heat bath
and $K_p$ gives the effective (inverse) temperature 
characterizing the randomness (see later).
The process I is relating to the ZFC.
At initial time $t=0$, 
the system is kept in the equilibrium state 
of a temperature $K_p$ with zero field;
hereafter, we use a terminology ``temperature" for $K$ and $K_p$ 
instead of ``inverse temperature". 
The temperature is immediately changed
(usually quenched) and the system relaxes 
in a heat bath with another temperature $K$ in $t>0$.
The average for dynamical ensembles in this process 
is denoted by $\langle\cdots\rangle_K^{K_p}$.
The process II is relating to the field quenching (FQ) 
\cite{NordEA87,Rieger93}.
The system starts from the all-up state 
${\bf F} =(+,+,\cdots ,+)$ at $t=0$ and 
relaxes in the same heat bath as in the process I in $t>0$;
the average is denoted by $\langle\cdots\rangle_K^{\bf F}$.
Since the all-up state provides the state in the strong-field limit,
this represents the process 
with the field quenched from $\infty$ to zero at $t=0$.
Note that the FQ is not equivalent to the FC 
in which the applied field is more weak 
and is quenched after the waiting time.
In the following, we show the derivation of 
Eq.\ (\ref{eq:AR}) which we call the {\it aging relation}
using the method of gauge transformation,
\cite{Nishim81,Kitata92,OzekiN93,Ozeki95}
and discuss the physical meaning of it.

Since randomness and frustration make it difficult to examine 
SG systems analytically as well as numerically,
only few things have been confirmed definitely.
The method of gauge transformation 
is a powerful technique to derive exact results 
in the $\pm J$ or the Gaussian Ising spin glasses irrespective of 
the dimensionality.
It provides the internal energy and an upper bound on the 
specific heat as non-singular functions of the temperature 
on a special line in the randomness-temperature phase diagram. 
This line is called the Nishimori line \cite{Nishim81}.
Further, it provides a plausible argument for the absence of 
re-entrant transition from the FM phase to non-FM one (SG in $3d$)
\cite{Kitata92}.
The method has been generalized to other random systems 
with various symmetries such as the XY gauge glass \cite{OzekiN93}.
Recently, it is extended to treat dynamical systems \cite{Ozeki95}.

The Hamiltonian we consider is
\begin{equation}
{\cal H} =J\widetilde{\cal H} ({\bf S}; \bbox{\omega})
=-J\sum_{\langle ij \rangle} \omega_{ij} S_i S_j ,
\end{equation}
where $S_i$ takes $\pm 1$, and 
${\bf S} =(S_1 ,S_2 ,\cdots ,S_N )$ represents a configuration of 
the total $N$ spins.
The set $\bbox{\omega} =(\omega_{12},\cdots )$ represents a configuration of 
the total $N_{\rm B}$ bonds, and
the summation is taken over all bonds; 
while we make no restrictions on the type or the dimension of the lattice, 
one may suppose usual nearest-neighbor 
interactions on the $d$-dimensional hypercubic lattice. 
For a particular bond configuration $\bbox{\omega}$,
the equilibrium distribution is defined by 
$\rho_{\rm eq} ({\bf S};K,{\bbox{\omega}})=
\exp\big\{-K\widetilde{\cal H}({\bf S};{\bbox{\omega}})\big\}
/Z(K,{\bbox{\omega}})$.
The exchange interaction $J_{ij} =J\omega_{ij}$ is a random variable. 
We treat both the $\pm J$ distribution and 
the Gaussian distribution with the variance unity.
The average for bond configurations is denoted by
$[\cdots ]_{\rm c}$.
The general form of the bond distribution 
in a gauge-symmetric model \cite{OzekiN93} is expressed as
\begin{equation}
P({\bbox{\omega}} ;K_p )={D(\bbox{\omega})\over Y(K_p )}
\exp\big\{-{K_p} \widetilde{\cal H} ({\bf F};{\bbox{\omega}})\big\},
\label{defP}
\end{equation}
where ${\bf F} =(+,+,\cdots ,+)$.
Functions
$\omega_{ij}$, $K_p$, $D(\bbox{\omega})$ and $Y(K_p )$
are summarized in table \ref{tab:s}.
In both distributions, $K_p$ controls the randomness;
${K_p} =0$ and $\infty$ correspond to 
the most random case and the non-random case, respectively.
The Nishimori line is located on $K=K_p$.

Since the Ising system has no intrinsic dynamics, 
we consider a Markov process for each bond configuration:
The density of state obeys the master equation 
\cite{VanKam81},
\begin{equation}
{{\rm d}\over{\rm d} t} \rho_t ({\bf S} )~=~
\sum_{{\bf S}'} \Gamma ({\bf S}|{\bf S}') \rho_t ({\bf S}' ).
\end{equation}
The solution of the master equation is formally given by 
\begin{equation}
\rho_t ({\bf S} )=\sum_{{\bf S}'} \,
\langle{\bf S}|{\rm e}^{t\bbox{\Gamma}}|{\bf S}' \rangle\,\rho_0 ({\bf S}' ).
\end{equation}
The matrix element 
$\langle{\bf S}|{\rm e}^{t\bbox{\Gamma}}|{\bf S}' \rangle$
plays a role of 
a Green's function for a time interval $t$.
The matrix $\bbox{\Gamma}$ is composed of non-negative off-diagonal elements,
and satisfies the detailed balance and the conservation of the probability.
We consider both the Metropolis \cite{MeRRTT53} and 
the Glauber dynamics \cite{Glaube63}. 
The detailed expressions of $\bbox{\Gamma}$ for these dynamics are 
shown in Ref.\ \cite{Ozeki95}.

The gauge transformations for functions of 
${\bf S}$ and ${\bbox{\omega}}$ are defined by
\begin{eqnarray}
U_{\bbox{\sigma}} ~~&:&~~ S_i        ~~\longrightarrow~ 
S_i \sigma_i ,\\
V_{\bbox{\sigma}} ~~&:&~~ \omega_{ij} ~\longrightarrow~ 
\omega_{ij} \sigma_i \sigma_j ,
\end{eqnarray}
where 
${\bbox{\sigma}} =(\sigma_1 ,\sigma_2 ,\cdots ,\sigma_N)$
is an arbitrary state of $N$ Ising spins.
The Hamiltonian 
is invariant under the transformation $U_{\bbox{\sigma}} V_{\bbox{\sigma}}$,
\begin{equation}
U_{\bbox{\sigma}} V_{\bbox{\sigma}}~
\widetilde{\cal H}({\bf S} ;{\bbox{\omega}} )~=~
\widetilde{\cal H} ({\bf S} ;{\bbox{\omega}} ).
\label{giH}
\end{equation}
Another important property in the static case is the invariance of summation
(or integral)
for ${\bf S}$ and ${\bbox{\omega}}$;
\begin{eqnarray}
\sum_{{\bf S}}\cdots ~&=&~\sum_{{\bf S}} U_{\bbox{\sigma}}\cdots ,
\label{invsU}\\
\sum_{\bbox{\omega}} \cdots ~&=&~
\sum_{\bbox{\omega}} V_{\bbox{\sigma}} \cdots .
\label{invsV}
\end{eqnarray}
Note that we use the terminology ``gauge invariant" only for
functions of the set ${\bbox{\omega}}$ invariant under $V_{\bbox{\sigma}}$.
The bond average of 
a gauge-invariant function $Q({\bbox{\omega}})$ can be expressed as
\cite{OzekiN93,Ozeki95}
\begin{equation}
\big[Q({\bbox{\omega}})\big]_{\rm c}
=\sum_{\bbox{\omega}} 
{D({\bbox{\omega}})Z({K_p} ,{\bbox{\omega}} ) \over 2^N Y({K_p} )} \,
Q({\bbox{\omega}}) .
\label{ragiQ}
\end{equation}
To treat dynamical systems, another transformation for ${\bf S}'$
is necessary; 
\begin{equation}
U'_{\bbox{\sigma}} ~~:~~ S'_i         ~\longrightarrow~ 
S'_i \sigma_i .
\end{equation}
We have shown the invariance of time evolution
\begin{equation}
U_{\bbox{\sigma}} U'_{\bbox{\sigma}} V_{\bbox{\sigma}}~
\langle{\bf S}|{\rm e}^{t\bbox{\Gamma}}|{\bf S}'\rangle
=\langle{\bf S}|{\rm e}^{t\bbox{\Gamma}}|{\bf S}'\rangle,
\label{giETW}
\end{equation}
for the Metropolis and the Glauber dynamics \cite{Ozeki95}.
This leads us to the relations obtained previously;
\begin{eqnarray}
\big[\langle S_i (t) \rangle^{\bf F}_K \big]_{\rm c}&=&
\big[\langle S_i (0) S_i (t) \rangle^{K_p}_K \big]_{\rm c},
\label{eq:SFSS}
\\ 
\big[\langle {\cal H} (t) \rangle^{\bf F}_K \big]_{\rm c}&=&
\big[\langle {\cal H} (t) \rangle^{K_p}_K \big]_{\rm c} .
\label{eq:HFH}
\end{eqnarray}
Equation (\ref{eq:SFSS}) is a generalization of 
the fluctuation-dissipation theorem in the region 
far from equilibrium. 

We examine the aging by analyzing
the waiting-time dependence of nonequilibrium auto-correlation functions
\cite{CugliK94,Rieger93}
for the processes I and II
instead of the remanent magnetization like in experiments.
Since the auto-correlation function is a fundamental quantity to exhibit
the dynamical structure in the system,
we believe it detects a typical aging feature of the process if any.
In the language of the master equation, they are expressed as
\begin{eqnarray}
\langle\!&S_i&\!(t_{\rm w}) S_i (t+t_{\rm w})\rangle_K^{\bf F}
\nonumber\\
&=&\sum_{{\bf S}_1 {\bf S}_2}
S_{2i}\langle{\bf S}_2|{\rm e}^{t\bbox{\Gamma}}|{\bf S}_1\rangle
S_{1i}\langle{\bf S}_1|{\rm e}^{t_{\rm w}\bbox{\Gamma}}|{\bf F}\rangle,\\
\langle\!&S_i&\!(t_{\rm w}) S_i (t+t_{\rm w})\rangle_K^{K_p}
\nonumber\\
&=&\!\!\!\sum_{{\bf S}_0 {\bf S}_1 {\bf S}_2}\!\!\!
S_{2i}\langle{\bf S}_2|{\rm e}^{t\bbox{\Gamma}}|{\bf S}_1\rangle
S_{1i}\langle{\bf S}_1|{\rm e}^{t_{\rm w}\bbox{\Gamma}}|{\bf S}_0\rangle
\rho_{\rm eq} ({\bf S}_0;K_p ,\bbox{\omega}).
\nonumber\\
\end{eqnarray}
{}From
the invariances (\ref{giH})-(\ref{invsV}) and (\ref{giETW}), 
it is easily seen that 
$\langle S_i(t_{\rm w}) S_i (t+t_{\rm w})\rangle_K^{\bf F}$
is transformed as
\begin{equation}
V_{\bbox{\sigma}}
\langle S_i(t_{\rm w}) S_i (t+t_{\rm w})\rangle_K^{\bf F} =
\langle S_i(t_{\rm w}) S_i (t+t_{\rm w})\rangle_K^{\bbox{\sigma}},
\label{eq:VSSSS}
\end{equation}
while $\langle S_i(t_{\rm w}) S_i (t+t_{\rm w})\rangle_K^{K_p}$
is gauge invariant; 
$\langle\cdots\rangle_K^{\bbox{\sigma}}$ expresses the average
for the process starting from a fixed state $\bbox{\sigma}$.
Note that the probability distribution (\ref{defP}) is transformed as
\begin{equation}
V_{\bbox{\sigma}} 
P({\bbox{\omega}};K_p )=
{D(\bbox{\omega}) Z(K_p, {\bbox{\omega}} )\over Y({K_p} )}
\,\rho_{\rm eq} ({\bbox{\sigma}};K_p ,\bbox{\omega}).
\label{gtP}
\end{equation}
Using Eqs.\ (\ref{ragiQ}), (\ref{eq:VSSSS}) and (\ref{gtP}),
we derive the relation (\ref{eq:AR})
for any waiting time $t_{\rm w}$, any time interval $t$,
any temperature $K$ and any degree of randomness $K_p$.
We call it the ``aging relation", since 
it relates aging phenomena in two distinct processes
whatever waiting-time dependence is essential for the aging.
Equation (\ref{eq:SFSS}) is a special case ($t_{\rm w}=0$) of it.

The aging relation contains 
parameters $K$ and $K_p$ characterizing the relaxation in $t>0$.
One can examine the physical meaning of it
for each phase by choosing $(K,K_p)$ appropriately.
A typical phase diagram in randomness-temperature plane 
for Ising spin glasses is shown in Fig.\ 2;
possible Griffiths phase and mixed phase are indicated.
Note that the initial temperature of the process I is always located 
on $K=K_p$.
In the SG phase where $K_p$ indicates a high temperature (small $K_p$),
the process I expresses the ZFC observed in experiments.
Therefore, the aging relation relates aging phenomena 
for the ZFC and the FQ processes.
The same relation was pointed out 
in real experiments \cite{NordEA87,LeflEA94}, in which
the waiting-time dependence of the remanent magnetization in the ZFC process
seemed equivalent with that in the FQ process.
Since the aging relation reveals just the equivalence 
of the gauge-invariant dynamical structure,
the amplitude of the magnetization which is not gauge invariant
does not have such relation,
while dynamical behavior itself may have it
as observed in these experiments.
At a glance, the aging relation expresses a trivial fact,
since both initial states are located in the PM phase and 
both systems relax into the SG phase.
However, the equivalence at any waiting time, 
which means the equivalence of the dynamical structure
at any stage of relaxation, is non-trivial. 
Further investigations in this direction would be helpful
to understand the dynamical structure of the system.

Next, let us consider the aging relation on $K=K_p$ (Nishimori line),
where the temperature keeps a constant at any time in the process I.
Then, the rhs of Eq.\ (\ref{eq:AR}) reveals 
the equilibrium auto-correlation function, 
which is independent of the waiting time $t_{\rm w}$:
\begin{equation}
\left[\langle S_i (t_{\rm w}) S_i (t+t_{\rm w})
\rangle_K^{\bf F}\right]_{\rm c} =
\left[\langle S_i (0) S_i (t)
\rangle_K^{\rm eq}\right]_{\rm c} .
\label{eq:AReq}
\end{equation}
$\langle\cdots\rangle_K^{\rm eq}$ denotes the dynamical average
for the equilibrium process.
Equation (\ref{eq:AReq}) is derived from the fact that 
the equilibrium distribution $\rho_{\rm eq} (K)$ 
is an eigenstate of $\bbox{\Gamma}$ with zero eigenvalue.
Therefore, the auto-correlation function in the FQ process
is independent of the waiting time
suggesting the absence of aging on the Nishimori line.
Further, it is noted that Eq.\ (\ref{eq:AReq}) provides 
efficient Monte Carlo calculations 
for equilibrium auto-correlation functions;
the equilibration can be omitted in the FQ process.

The aging has been considered an inherent property in the SG phase
from experimental \cite{NordEA87,VincHO92,LeflEA94}
and theoretical \cite{KoperH88,FisheH88,SibanH89,Rieger93} view points.
Since the aging phenomenon
is a typical realization of the complex phase space for slow dynamics,
it could occur in other complex phases.
In such a phase, the aging would also be inherent, which means that 
it occurs whole in the phase when it is observed in some part of the phase.
It has been pointed out that there exists a dynamically singular phase
called the Griffiths phase
\cite{Takaya95,RandSP85,Ogiels85,TakanM95,KomoHT95}
between the critical temperature
of the pure system and the phase boundaries of low temperature phases 
(the FM or the SG) --- see Fig.\ 2.
Concerning to the Griffiths phase,
two cases can be considered to understand the above result
if the aging is an inherent property.
(a) {\it There is no aging whole in the Griffiths phase},
if the Nishimori line intersects the Griffiths phase as in Fig.\ 2.
Then, even if slow dynamics is observed in the Griffiths phase,
it is quite different from that in the SG phase
\cite{DerriW87}.
However, the region of the Griffiths phase has not yet been 
determined definitely:
Figure 2 was proposed just following an analogy 
with the dilute ferromagnet \cite{Griffi69}.
One can also consider that 
(b) {\it there is no Griffiths phase at least around the Nishimori line.}
This allows both the existence of the Griffiths phase below $K=K_p$
and the absence of it.
Another possibility is that 
the aging is not inherent at least in the Griffiths phase.
Although it is not obvious which is correct in the present framework,
the above result restricts the aging phenomenon
and the region of the Griffiths phase in future investigations.

Following the results of the SK model and experiments \cite{BindeY86},
one should consider the possibility of 
the mixed phase between the FM and SG phases 
-- see Fig.\ 2.
It is reasonable to consider that the aging occurs in the mixed phase, 
since the SG feature in the mixed phase provides
typical slow dynamics which reveals the aging.
Let us consider the temperature below the multicritical point on $K=K_p$, 
where the spontaneous magnetization appears.
In such a region,
it is not clear that the FQ process is appropriate 
for the observation of the aging feature,
since the initial all-up state is not so different 
from the final equilibrium state with broken symmetry.
As seen above, the amplitude of the magnetization
is not important in the aging phenomenon.
While the up-down symmetry is broken in the FQ process,
the dynamical behavior is equivalent with that in the equilibrium 
(symmetric) process --- Eq.\ (\ref{eq:AReq}).
Thus, we assume that the FQ process provides an aging feature 
even in such a symmetry-broken region.
If the aging is inherent in the mixed phase,
the relation (\ref{eq:AReq}) indicates that 
the Nishimori line does not enter the mixed phase.
This restricts the topology of the phase diagram,
and is consistent with the result of the SK model.
%This argument is important, 
%since it has been difficult to discuss the mixed phase 
%theoretically in short range models.

In summary, we derive the relation (\ref{eq:AR}) 
on the aging phenomena for two nonequilibrium processes.
This provides some restrictions to the Griffiths and the mixed phases.
Similar relations can be derived for other gauge-symmetric quantities
such as the SG susceptibility,
and in other processes like in the simulated annealing.
The present theory is applicable to various systems:
Any gauge-symmetric distribution for randomness 
instead of the Gaussian and the $\pm J$,
any dynamics which satisfies Eq.\ (\ref{giETW})
instead of the Glauber and the Metropolis,
any dimensionality and lattice ({\it e.g.} the SK model).
It can be extended to other gauge-symmetric systems such as 
the XY gauge glass \cite{OzekiN93};
${\cal H} = -J\sum_{\langle ij\rangle} 
\cos(\phi_i -\phi_j +A_{ij})$.
It has been pointed out that granular systems of 
type-II superconductors in a magnetic field are expressed this model.
There exists a similar relation \cite{Ozeki96} in such materials.

The author would like to thank 
H. Rieger, E. Vincent, J. M. Hamman, I. A. Campbell, S. Miyashita,
K. Nemoto, J. Kurchan and L. F. Cugliandolo
for valuable discussions and comments.

\begin{center}
\begin{figure}
\psfig{file=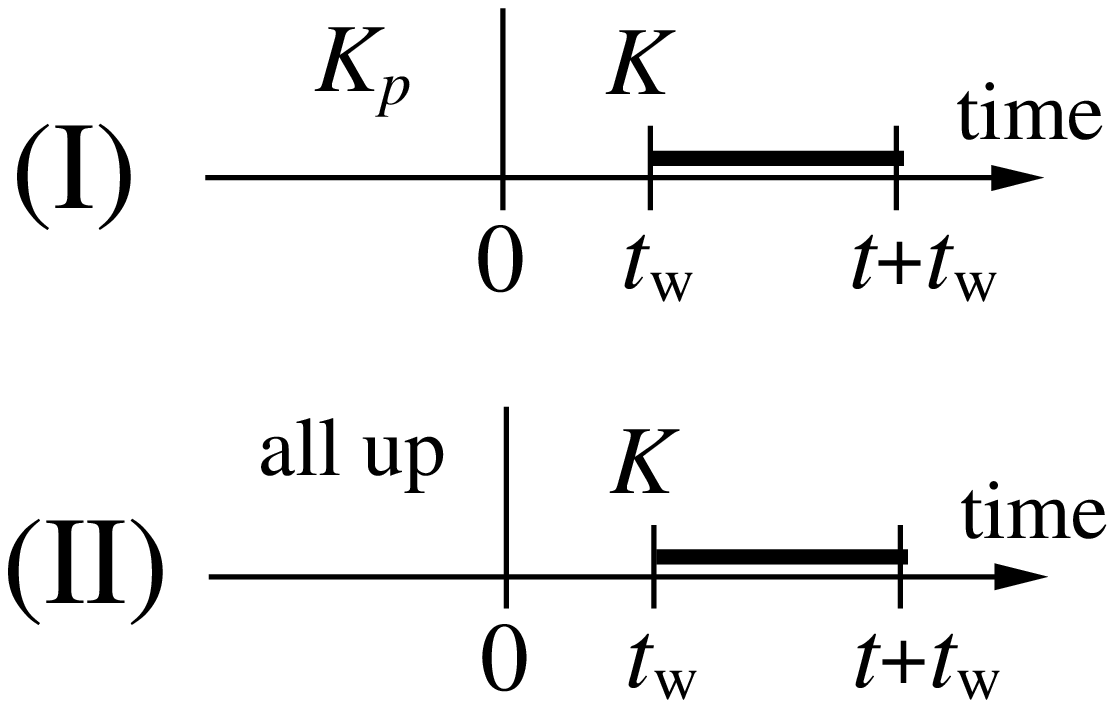,scale=0.45}
\caption{
Illustrations of two nonequilibrium processes, I and II.
Between $t_{\rm w}$ and $t+t_{\rm w}$, the correlation is measured.
\label{fig1}}
\end{figure}
\begin{figure}
\psfig{file=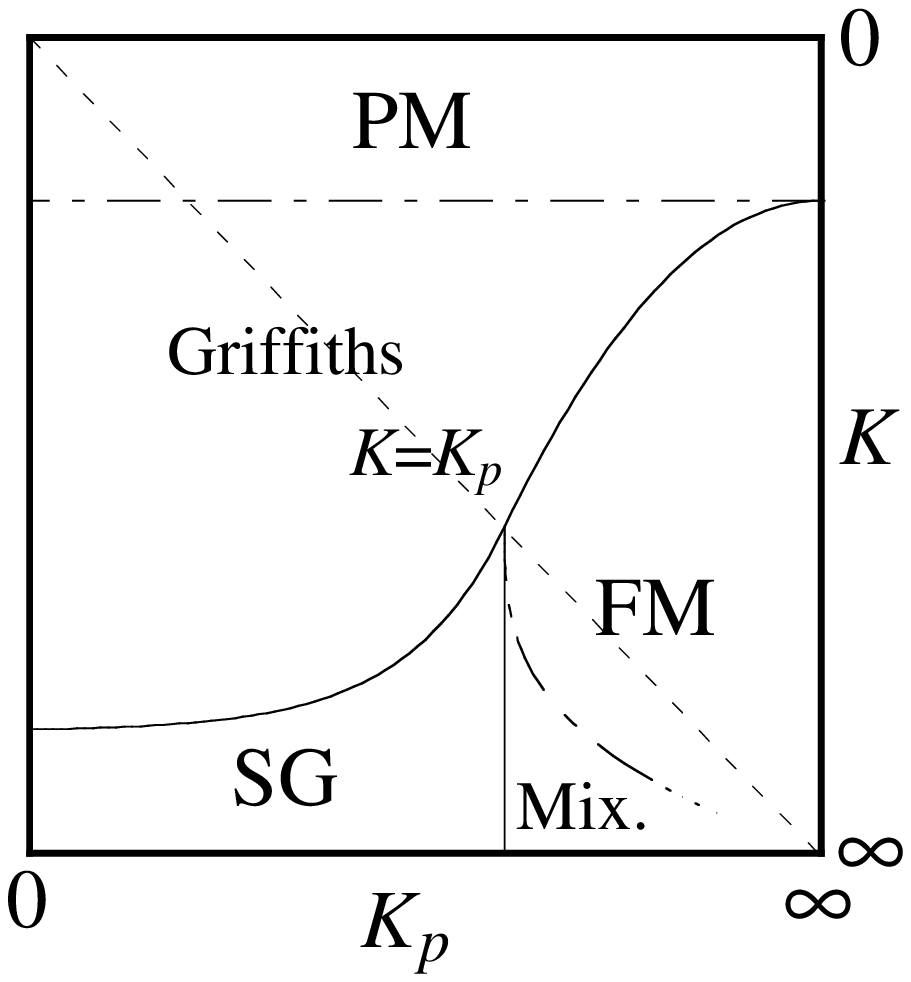,scale=0.48}
\caption{
Typical phase diagram of Ising spin glasses 
in the ${K_p}$--$K$ plane.
The dashed line ($K=K_p$) is the Nishimori line.
Possible Griffiths phase and mixed phase (Mix) are indicated.
\label{fig2}}
\end{figure}
\end{center}
\begin{table}[h]
\caption{The summary of variables and functions
appeared in the bond distributions.
$N_{\rm B}$ is the number of bonds.
$p$ is the concentration of $+J$ bonds in the $\pm J$ distribution,
and $J_0$ is the center of distribution 
in the Gaussian distribution with the variance unity.
}
\begin{tabular}{l|cc}
& $\pm J$ & Gaussian \\
\hline
$\omega_{ij}$ & $\pm 1$ & any real values \\
$K_p$ & ${1\over 2}\ln (p/1-p)$ & $J_0$ \\
$Y(K_p )$ & $(2\cosh{K_p} )^{N_{\rm B}}$ 
& $\exp\big({1\over 2} N_{\rm B}{K_p}^2 \big)$ \\
$D(\bbox{\omega})$ & $1$ 
& $\exp \big(-{1\over 2}\sum_{\langle ij\rangle} \omega_{ij}^2\big)$
\end{tabular}
\label{tab:s}
\end{table}

\begin{references}
\bibitem{Ozeki96}
Since no ordering field exists in superconducting systems,
the all-up state is prepared by the ordered state 
in the pure system at sufficiently low temperature.
In this literature, the parameter $K_p$ controls the magnetic field.
\end{references}
\end{document}